\documentclass[prl,nofootinbib,twocolumn]{revtex4}
\usepackage{bm,bbm}
\usepackage{amsmath,amssymb,amsthm,amsfonts}
\usepackage{mathrsfs}
\usepackage{ucs}
\usepackage[utf8x]{inputenc}
\usepackage{hyperref}
\usepackage{hepunits}

\begin{document}

\def\be{\begin{equation}}
\def\ee{\end{equation}}
\def\bed{\begin{description}}
\def\eed{\end{description}}

\def\bea{\begin{eqnarray}}
\def\eea{\end{eqnarray}}

\def\noi{\noindent}
\def\ba{\begin{array}}
\def\ea{\end{array}}

\def\half{\frac{1}{2}\,}
\def\third{\frac{1}{3}\,}
\def\quart{\frac{1}{4}\,}

\def\atH0{|_{H_{0}}}
\def\AtH0{\bigg|_{H_{0}}}
\def\barH{{\overline{H}}}
\def\barL{{\overline{L}}}
\def\barF{{\overline{F}}}
\def\barPhi{{\overline{\Phi}}}
\def\barG{{\overline{G}}}
\def\barW{{\overline{W}}}
\def\hatW{{\widehat{W}}}
\def\hatK{{\widehat{K}}}
\def\hatG{{\widehat{G}}}
\def\hatS{{\widehat{S}}}
\def\hatD{{\widehat{D}}}
\def\hatV{{\widehat{V}}}
\def\hatMD{{\widehat{\mathcal{D}}}}
\def\dslash{ ~\partial \hskip -0.17cm \slash \hskip+0.05cm }
\def\dd{\mathrm{d}}
\def\Re{\mathrm{Re}}
\def\Im{\mathrm{Im}}

\def\AA{{\bf (AA) \ }} 
\def\SH{{\bf (SH) \ }} 

\def\hd{\hat\delta}

\bibliographystyle{h-physrev}

\title{Consistent Decoupling of Heavy Scalars and Moduli in $\mathcal{N}=1$ Supergravity}

\author{Ana Achúcarro$^{1,2}$, Sjoerd Hardeman$^1$ and  Kepa Sousa$^1$}
\affiliation
{$^1$Instituut-Lorentz for Theoretical Physics, Leiden, The Netherlands, \\
{$^2$Department of Theoretical Physics, 
The University of the Basque Country UPV-EHU, 48940 Bilbao, Spain}}

\begin{abstract}
  We consider the conditions for integrating out heavy chiral fields
  and moduli in $\mathcal{N}=1$ supergravity, subject to two
  \emph{explicit} requirements. First, the expectation values of the
  heavy fields should be unaffected by low energy phenomena. Second,
  the low energy effective action should be described by
  $\mathcal{N}=1$ supergravity. This leads to a working definition of
  decoupling in $\mathcal{N}=1$ supergravity that is different from
  the usual condition of gravitational strength couplings between
  sectors, and that is the relevant one for inflation with moduli
  stabilization, where some light fields (the inflaton) can have long
  excursions in field space. It is also important for finding de
  Sitter vacua in flux compactifications such as LARGE volume and
  KKLT scenarios, since failure of the decoupling condition
  invalidates the implicit assumption that the stabilization and
  uplifting potentials have a low energy supergravity description.

  We derive a sufficient condition for supersymmetric decoupling,
  namely, that the Kähler invariant function $G = K + \log |W|^{2}$ is
  of the form $G = L(\mathrm{light}, H(\mathrm{heavy}))$ with $H$ and
  $L$ arbitrary functions, which includes the particular case $G =
  L(\mathrm{light}) + H(\mathrm{heavy})$. The consistency condition
  does \emph{not} hold in general for the ansatz $K =
  K(\mathrm{light}) + K(\mathrm{heavy})$, $W = W(\mathrm{light}) +
  W(\mathrm{heavy})$ and we discuss under what circumstances it does
  hold.
\end{abstract}

\maketitle

The viability of theories based on extra dimensions, in particular
string theory, relies on being able to stabilize and integrate out the
fields (moduli) that describe the shapes and sizes of those extra
dimensions, for which so far there is no observational evidence. In
flux compactifications \cite{Giddings:2001yu} some moduli are
stabilized at a high energy scale and decouple from the low energy
theory. From that moment on we never see them in the effective low
energy description.

Unlike in global supersymmetry, complete decoupling is of course
impossible in supergravity --even in principle-- because gravity
couples to all fields; so at low energies one is usually satisfied
with gravitational strength couplings between the heavy, stabilized,
fields and the low energy fields.  However such interaction terms are
of order $O(G_{\mathrm{Newton}} E^{2}) = O(E^2/M_{P}^{2})$, where $E$
is the energy scale and $M_{P} \approx 2.4 \times 10^{18} \GeV$ the
reduced Planck mass.  Even if they are strongly suppressed at low
energy and in particle accelerators, these couplings become sizeable
at the energy scales relevant to the early Universe, and one must look
for a more robust definition of decoupling that can be extrapolated
over a wide range of energy scales. The purpose of this note is to
provide such a definition, and a simple test of whether it holds in
specific models.

There are at least two situations in which the details of decoupling
are important. One is supersymmetry breaking, which will affect the
heavy fields in a way that is not accounted for in the low energy
effective action. Uplifting in KKLT scenarios \cite{Kachru:2003aw} is
a prime example. The second is inflation with moduli stabilization,
because the inflaton, which is a low energy field in this language,
can have its expectation value vary over many Planck-masses.

Here we take a bottom-up approach and try to find for what types of
Supergravity couplings we can be sure that the heavy moduli will not
shift from their expectation values due to low energy processes. We do
not require small gravitational coupling to the light(er) fields
because instead we rely on supersymmetry to partially protect the
expectation values of the heavy moduli.

It must be stressed that what we are proposing here, building on
arguments by other authors
\cite{Choi:2005ge,deAlwis:2005tg,deAlwis:2005tf,Binetruy:2004hh,Achucarro:2007qa},
is a simple consistency test. It checks explicitly what is implicitly
assumed by the very use of a low energy effective action. So it is
somewhat surprising to find that the most common ansatz for decoupled
fields in the literature, the standard ``gravitational strength
coupling'' ansatz, generically fails the test. It partly explains the
difficulties encountered in supergravity models of inflation with
moduli stabilization. The problem essentially disappears for
consistently decoupled moduli (see
\cite{Davis:2008sa,Achucarro:2007qa}).

\section{Notation and conventions}

We will use units in which $M_{P} = 1$. We start by recalling that the
$\mathcal{N}=1$ supergravity action involving scalars and gauge fields
(chiral and gauge superfields)
\begin{equation}
  S = \int d^{4}x\sqrt{-g} (\frac{1}{2}R + T - V + \mathcal{L}_{\mathrm{gauge}}) \label{Sdef}
\end{equation}
is entirely described by three functions of the scalars: the Kähler
potential $K(z,\bar{z})$, the holomorphic superpotential $W(z)$ and
the gauge kinetic functions $f_{ab}(z)$. The action and the
supersymmetry transformations are invariant under Kähler
transformations,
\begin{equation} \label{kahlertrans}
 K \to K + h(z) + {\bar h}(\bar{z}) \qquad W \to W e^{-h(z)},
\end{equation}
with $h(z)$ an arbitrary holomorphic function. Actually, if $W \neq
0$, they only depend on the Kähler invariant function $G = K +\log
|W|^2$ and the gauge kinetic functions.

In terms of these functions, the different terms of the action (eq.
\ref{Sdef}) are
\begin{subequations}
 \begin{align}
  T &= G_{i\bar{\jmath}}\mathscr{D}_{\mu}z^{i}\mathscr{D}^{\mu}\bar{z}^{\bar{\jmath}} \label{Skinetic}\\
  G_{i\bar{\jmath}} &\equiv \partial_{i} \partial_{\bar{\jmath}} G = \partial_{i} \partial_{\bar{\jmath}} K,
 \end{align}
\end{subequations}
where $\mathscr{D}_{\mu}z^{i} = \partial_{\mu}z^{i} -
W_{\mu}^{a}\eta_{a}^{\phantom{a}i}(z)$, with
$\eta_{a}^{\phantom{a}i}(z)$ the killing vector that defines the gauge
transformations of the scalars,
$\delta_{a}z^{i}=\eta_{a}^{\phantom{a}i}(z)\alpha^{a}$, where
$\alpha^{a}$ is the gauge parameter. The scalar potential includes a
contribution from F-terms and D-terms
\begin{equation} 
 V =  V_{F} + V_{D},
\end{equation}
where $V_{F}$ and $V_{D}$ are
\begin{subequations}
  \begin{align}
    V_{F} &=  G_{i\bar{\jmath}}F^{i}F^{\bar{\jmath}}  - 3e^{G} \label{scalarf}\\
    V_{D} &= \frac{1}{2} \Re(f_{ab}){D}^{a}D^{b} \label{scalard}.
  \end{align}
\end{subequations}
The $F^{i}$ and $D^{a}$ are the auxiliary fields of the chiral and gauge superfields respectively. They have equations of motion that can be solved algebraically in terms of the chiral fields 
\begin{subequations}
 \begin{align}
   F^{i} &= e^{G/2}G^{i\bar{\jmath}} {G}_{\bar{\jmath}}\label{fterms} \\
   D^{b} &= i(\Re f_{ab})^{-1} \eta_{a}^{\phantom{a}i} G_{i}.
 \end{align}
\end{subequations}
We are assuming that there are no constant Fayet-Iliopoulos terms present; these require a more careful treatment that will be given elsewhere.

The (holomorphic) gauge kinetic functions $f_{ab}(z)$ determine the kinetic terms of the gauge fields.
\begin{equation}
 \begin{split} 
   \mathcal{L}_{\mathrm{gauge}} = &- \frac{1}{4}(\Re f_{ab})F^{a}_{\phantom{a}\mu\nu}F^{b\mu\nu} + \\
   & \frac{1}{4\sqrt{-g}} (\Im
   f_{ab})F^{a}_{\phantom{a}\mu\nu}\epsilon^{\mu\nu\rho\sigma}F^{b}_{\phantom{b}\rho\sigma}.
 \end{split}
\end{equation}

\section{Consistent decoupling of scalar fields in $\mathcal{N}=1$
  supergravity}
In what follows we consider two sets of fields, heavy ($H$) and light
($L$), and assume the heavy fields are stabilized at an expectation
value $H=H_{0}$, an extremum of the scalar potential for the heavy
moduli. If the heavy field is a singlet under all low energy
symmetries and its mass is large enough it will decouple from low
energy phenomena and can be integrated out, leaving an effective
theory for the light degrees of freedom. To make this distinction, we
will from now on use hatted quantities to indicate the full theory,
including heavy and light fields, and unhatted quantities for the
effective theory involving light fields only
\begin{equation}
 S (L, \barL) = \hatS (H_{0}, \barH_{0}, L, \barL).
\end{equation}

We are interested in the case in which the resulting effective theory
is also described by $\mathcal{N}=1$ supergravity. In this case, there
should be an effective $K$ and $W$ (or $G$) depending only on the
light fields, from which to compute the low energy action $S$ and
supersymmetry transformations
\begin{align}
 G[L,\barL] &= \hatG[L,\barL,H_{0},\barH_{0}] \\
 \delta_{\epsilon} L = \hat{\delta}_{\epsilon} L\atH0 &= f[L,G(L,\barL)] \\
 \hat{\delta}_{\epsilon} H\atH0 &= 0.
\end{align}
Notice that the F-terms (eq. \ref{fterms}) of the heavy fields must
vanish because the supersymmetry transformations read,
\begin{equation}
 \hat{\delta}_{\epsilon} H \sim \chi \epsilon, \qquad \hat{\delta}_{\epsilon} \chi \sim \dslash H \epsilon  -\frac{1}{2} F \epsilon
\end{equation}
and if the F-terms are non-zero a supersymmetry transformation will
generate light fermions that are not in the low energy effective
action. Thus, the heavy fields cannot contribute to supersymmetry
breaking, leading to
\begin{equation}\label{MDHW}
 \partial_{H}\hatG\atH0=0 \qquad \textrm{or} \qquad \hatMD_{H}\hatW\atH0 = 0,
\end{equation}
(see also \cite{deAlwis:2005tg}) where
$\hatMD_{i}\hatW=\partial_{i}\hatW+(\partial_{i}\hatK)\hatW$ is the
Kähler covariant derivative that transforms as $\hatMD_{i}\hatW \to
e^{-h(z)} \hatMD_{i}\hatW$ under Kähler transformations. Note that
$\hatMD_{H}\hatW=0$ is the condition used in flux compactifications
\cite{Giddings:2001yu} and by extension in KKLT \cite{Kachru:2003aw}
and LARGE volume scenarios \cite{Balasubramanian:2005zx}, where the
complex structure moduli are stabilized at a supersymmetric point
before uplifting.

The Kähler metric should be block diagonal in the light and heavy fields when evaluated at $H_{0}$, otherwise propagators will mix these two sets of fields.  Additionally, the truncation $H=H_{0}$ must of course be a consistent truncation. This means that the equations of motion of the light fields derived from the effective theory are the same as the equations of motion obtained from the full theory. To zeroth order in the fluctuations of the heavy fields:
\begin{equation}\label{decouple}
\frac{\delta \hatS}{\delta L}\AtH0 = \frac{\delta\hatS\atH0}{\delta L} = \frac{\delta S }{\delta L},
\end{equation}
ensuring that the fluctuations of $H$ are not sourced by the light
fields.  In particular, the heavy fields should be singlets under the
surviving gauge group at low energies (otherwise they remain coupled
to the light fields by the gauge interaction).  In what follows we
will consider $f_{ab} $ independent of the heavy fields. In that case
they do not contribute to the D-terms, which will only involve light
fields.

\section{Analysis of the consistency conditions}
The heavy fields thus need to be stabilized at an expectation value $H_{0}$, where $H_{0}$ is the solution to eq. (\ref{MDHW})
\be \left[\partial_{H} \hatW (H,L) + \partial_{H} \hatK(H, \barH, L, \barL)
\hatW(H,L) \right]\AtH0 = 0.
\label{DW}\ee
which implies $\partial_{H} \hatV\atH0=0$. The LHS is some function of
both the heavy and the light fields, let us call it $\Phi(H, \barH, L,
\barL)$. In general, the condition $\Phi=0$ (together with its complex
conjugate $\barPhi = 0$) relate the heavy and light fields. If we can
solve for $H$ we obtain an expression of $H_{0}$ as a function of the
light fields,
\begin{equation}
  H = H_{0} (L, \barL),
\end{equation}
which can be substituted back into $\hatK, \hatW$ to give an effective
action for the light fields
\begin{equation}
 S (L, \barL) = \hatS (H_{0}(L, \barL), \barH_{0}(L,\barL),L,\barL).
\end{equation} 

An immediate concern with the consistency of this procedure, pointed
out in \cite{deAlwis:2005tg}, is that in general this leads to a
non-holomorphic expression for the would-be effective superpotential
$W = {\hatW} (H_{0}(L,\barL), L) $.  However, this problem is easily
avoided: it does not arise if $\hatW$ is independent of $H$. The case
$\hatW = 0$ is obvious, so consider $\hatW \neq 0$. It is always
possible to perform a Kähler transformation that makes $\hatW$
constant
\begin{subequations}\label{kahlergauge}
 \begin{align}
  \hatW &\to 1 \\
  \hatK &\to \hatK + \log\hatW + \log\widehat{\barW} = \hatG \ .
 \end{align}
\end{subequations} 
In this so called Kähler gauge, eq. (\ref{DW}) reads 
\begin{equation}\label{dG} 
 \partial_{H} \hatG (H,\barH,L,\barL) = 0,
\end{equation} 
from which we can extract $H = H_{0}(L, \barL)$ and make the previous substitution directly into the Kähler invariant function without any inconsistency
(see also \cite{Curio:2006ea}):
\begin{equation}
  G = \hatG (H_{0}(L,\barL), \barH_{0}(L,\barL), L, \barL).
\end{equation}

In fact, the issue is not whether $H_{0}(L, \barL)$ is holomorphic but
rather whether it is \emph{a (non-trivial) function at all}. The
assumption that the heavy fields are {\it stabilized} at $H=H_{0}$ is
simply the condition that $H_{0}(L,\barL)= \mathrm{constant}$. Any
other dependence on the light moduli would translate into a constraint
on the light fields which would have to be accounted for explicitly in
the low energy action \cite{deAlwis:2005tf}. This is what we have to
avoid.

To summarize: the (rather obvious) mathematical condition for the
heavy fields to be integrated out consistently with an expectation
value $H_{0}$ and to decouple from the low energy fields is that the
system of equations
\begin{equation}\label{trunc}
  \partial_{H} \hatG \equiv \Phi(H,\barH,L,\barL) = 0,
\end{equation}
which is the same as (\ref{DW}) defined in the Kähler gauge (eq. \ref{kahlergauge}), admits the constant solution 
\begin{equation}
 H = H_{0}(L,\barL) = \mathrm{const}\qquad \barH = \barH_{0}(L,\barL) = \mathrm{const}.
\end{equation}
In spite of being obvious, this condition is not empty. For instance,
we will see below that it \emph{fails} generically for standard
couplings of the form $K=K_{1}+K_{2}$ and $W=W_{1}+W_{2}$.  But let us
first consider two specific situations in which the decoupling
condition does hold.
\begin{enumerate}
\item The consistency condition is trivially satisfied if the function
  $\Phi(H,\barH,L,\barL)$ has no explicit dependence on the light
  fields.  In this case integrating eq. (\ref{trunc}) recovers the
  condition found in \cite{Binetruy:2004hh}
  \begin{equation}\label{Gsum}
 \phantom{l}\quad\ \ \partial_{H} \hatG = \Phi(H,\barH) \to \hatG = \hatG_{1}(H, \barH) + \hatG_{2} (L, \barL)
\end{equation} 
and it is obvious that the Kähler metric is block diagonal in this
case. This ansatz has a long history \cite{Cremmer:1982vy} and allows
a detailed stability analysis of the heavy fields
\cite{Achucarro:2007qa,us}, in particular in the context of F-term
uplifting of flux compactifications.
\item On the other hand, this requirement is too restrictive. It is
  sufficient if the function $\Phi(H,\barH,L,\barL)$ factorizes:
  \begin{equation}
 \phantom{l}\quad\ \ \Phi(H,\barH, L, \barL ) = \Phi_{1}(H, \barH, L, \barL )~ \Phi_{2}(H,\barH) = 0
\end{equation}
in which case we just solve $\Phi_2 = \barPhi_{2} = 0$ to get constant
$H_{0},\barH_{0}$.  We cannot give the general form of $\hatG$ for
which this factorization occurs, but it will certainly hold if $\hatG$
has the following functional form:
\begin{equation}\label{mainresult}
 \hatG = f(L, \barL,g(H, \barH)) 
\end{equation}
since in that case eq. (\ref{MDHW}) is replaced by
\begin{equation}\label{Gofg}
 \partial_{H} g(H, \barH) = 0.
\end{equation}
\end{enumerate}

The first situation, eq. (\ref{Gsum}), is a special case of eq.
(\ref{Gofg}), with $\Phi_{1}$ constant. In both cases, the same
condition that makes $\hatG_{H}\atH0 = 0$ also implies that the Kähler
metric and the Hessian of V are block diagonal for any $\Phi_{1}$.
Indeed, from equation (\ref{Gofg}) we find that
\begin{equation}
\hatG_{LH}\atH0 = \partial_{L}\partial_{g} f(L, \barL, g(H, \barH)) \partial_{H} g(H, \barH)\atH0 = 0
\end{equation}
and further all mixed derivatives with only one derivative with
respect to the heavy field vanish. As $V_{LH}$ always contains terms
$\propto \hatG_{H}$ or $\propto (\partial_{L})^{n} \hatG_{H}$, which
vanish at $H_0$, the Hessian of $V$ is block diagonal.
\footnote{Note that it is always possible to diagonalize the Kähler metric or the Hessian of V at one point, but it is not necessarily the case that both diagonalizations are compatible, as we have here.}

\section{Consistent decoupling versus standard gravitational
  couplings}
Finally, we stress that the condition derived here has no direct
relation to the condition usually associated with gravitational
strength coupling. In fact, the ansatz
\begin{subequations}\label{gravcoupl}
 \begin{align}
  \hatK &= K_1(H,\barH) + K_2 (L, \bar L) \\
  \hatW &= W_1(H) + W_2 (L) \label{sepsuperpot}
 \end{align}
\end{subequations}
does not satisfy the decoupling condition in general. Suppose eq.
(\ref{MDHW}) admits a constant solution $H = H_{0}$. Then
\begin{equation} 
  0 = \partial_{H} W_{1} \atH0 + \partial_{H} K_{1} \atH0 [ W_{1}(H_{0}) + W_{2}(L) ],
\end{equation}
which only holds if
\begin{align}
  \partial_{H} K_{1} \atH0 &= 0 \ \Rightarrow \ \partial_{H} W_{1} \atH0 = 0 \nonumber \\
  \partial_{H} K_{1} \atH0 &\neq 0 \ \Rightarrow \ W_{2}(L) = -\frac{\partial_{H} W_{1} \atH0}{\partial_{H} K_{1} \atH0} - W_{1}(H_{0}) \nonumber \\
& \phantom {\neq 0 \ \Rightarrow \ W_{2}(L)} \ = \ \text{const.}
\end{align}

Another way to see this: since $\hatMD_{H}\hatW = 0$ does not
factorize, the (Kähler-gauge covariant) requirement that
it is independent of the light fields is (see also
\cite{BenDayan:2008dv})
\begin{equation}\label{vardh}
 \hatMD_{L}(\hatMD_{H}\hatW) = 0.
\end{equation}
Inserting the ansatz (eq.  \ref{gravcoupl}) then gives
\begin{equation}
 \partial_{H}K_{1}\atH0\partial_{L}W_{2} = 0.
\end{equation}
Unless $K_1(H, \barH)$ has no linear terms or $W_2(L) =
\textrm{constant}$, the condition will not be met. However, if
$W_{2}(L) = \textrm{constant}$ (e.g. no scale models
\cite{Cremmer:1983bf,Giddings:2001yu}) then equation (\ref{Gsum})
holds and $\hatW$ is trivially a product. On the other hand, we can
always expand $K_{1}(H,\barH)$ around $H_{0}$ and remove the linear
terms by a Kähler transformation (eq. \ref{kahlertrans}), but this
spoils the separability of the superpotential (eq. \ref{sepsuperpot}).

In other words, if two sets of fields have separable Kähler functions
$K = K_1(\mathrm{heavy}) + K_2(\mathrm{light})$, the addition of their
superpotentials does not respect the decoupling condition except in
special cases (and, incidentally, neither does it guarantee
gravitational strength couplings if $K_1(heavy) = O(M_p^2)$, as is
usual for moduli).

\section{Discussion}
In this Letter we have studied how to integrate out heavy scalars and
moduli and their superpartners in $\mathcal{N}=1$ supergravity,
subject to two \emph{explicit} requirements. First, the expectation
values of the heavy fields should be unaffected by low energy
phenomena, in particular supersymmetry breaking. Second, the low
energy effective action should be described by $\mathcal{N}=1$
supergravity. This is what we call \emph{consistent decoupling}.

If the heavy fields are stabilized at a critical point of the
potential, integration of the whole superfield requires that the
F-terms should be zero \cite{Binetruy:2004hh}. The criterion for
consistent decoupling is that the expectation value of the heavy
scalars $H$ should not depend on the light fields $L$
\cite{deAlwis:2005tf}. Our main result is a class of Kähler invariant
functions that satisfy the condition, given in eq (\ref{mainresult}):
\[\hatG = f(L, \barL,g(H, \barH)).\]

This functional form guarantees that the Kähler metric and Hessian of
$V$ are simultaneously block diagonal in the heavy and light fields.
It also allows the embedding of BPS solutions of the low energy
effective theory into the full theory without destroying their BPS
character (if the F-terms of the heavy fields are zero and in the
absence of constant Fayet-Iliopoulos terms, the supersymmetric
transformation of the gravitino depends only on the light fields). We
would expect the BPS character to survive quantum corrections -now in
the full theory-.  So at least in this special case it would seem
possible to ``screen'' the heavy, decoupled fields from the effects of
(partial) supersymmetry breaking in the low energy sector.

We only have experimental access to $G$, the effective low energy
theory, and there is a large class of supergravity models (read a
landscape of compactifications), characterized by $\hatG$, in which
the low energy theory could be embedded. Here, $\hatG$ includes all
stringy, perturbative and non-perturbative effects. The decoupling
condition restricts the allowed functional form of $\hatG$ and
therefore the class of models that are consistent with the assumption
of decoupling that is implicit in our use of $G$. From the point of
view of model building, it provides a simple test that has not been
considered before. There are string compactifications which
approximately satisfy the decoupling condition in the form
(\ref{Gsum}), such as some LARGE volume scenarios (LVS)
\cite{Balasubramanian:2005zx,Conlon:2005ki,Conlon:2006wz,Conlon:2007dw}.

To see this, note first of all that the tree level or GKP limit
\cite{Giddings:2001yu} of $\hatG$ satisfies eq. (\ref{Gsum}) with the
complex structure moduli and the dilaton $S$ playing the role of the heavy
fields.
Assume the usual form for the leading non-perturbative and $\alpha'$
corrections, $\hatW = W_{\text{GKP}}(H) + W_{\text{np}}(L)$, 
$\delta \hatK \sim 2 (S+\bar S)^{3/2} /\text{vol}$.
Ignoring for a moment the dilaton dependence of $\delta \hatK$, we find
for the complex structure moduli
\begin{equation}
  \partial_H \hatG = \partial_H K_{\text{heavy}}(H) 
  + \frac{\partial_H W_{\text{GKP}} (H)}   {W_{\text{GKP}} (H)}    
    \biggl[ 1 + \delta(L,H) \biggr]^{-1} \ ,
\end{equation}
where $\delta = W_{\text{np}} (L) / W_{\text{GKP}} (H)$. Including
dilaton effects adds a correction $\delta \sim (S+\bar S)^{3/2}
/\text{vol}$ (whichever is larger).  The condition of consistent
decoupling is violated by the L-dependence of $\delta$.  It is
negligible, $\delta \sim O(10^{-10})$, for an LVS vacuum with
parameters $A \sim 1$, $W_{\text{GKP}} (H_0) \sim 10$, $\text{vol}
\sim 10^{10}$, $A e^{-a_4 \tau_4} \sim 1 / \text{vol}$ (see
\cite{Balasubramanian:2005zx}). $^{2}$  In the mirror mediation scenarios
\cite{Conlon:2007dw} $\delta$ is even smaller.  By constrast, $\delta
\sim O(1)$ in a KKLT vacuum with parameters $A \sim O(1)$,
$W_{\text{GKP}}(H_0) \sim O(10^{-4}), aL \sim O(10)$ (see
\cite{Kachru:2003aw})).

Finally, we emphasize that the condition (eq. \ref{mainresult}) is not
easily expressed in terms of $K$ and $W$, in particular it has nothing
to do with gravitational strength couplings.  When $K=
K_1(\mathrm{heavy}) + K_2(\mathrm{light})$, the addition of
superpotentials does not lead to consistent decoupling in general
(whereas the product always does).  The problem considered here
illustrates once again the dangers of extrapolating our low energy,
weak gravity intuition, based on $K$ and $W$, to the very high energy
regimes encountered in the early Universe. Inflation model building is
hard enough as it is without these unnecessary complications.

\bigskip

\acknowledgments{ We thank Joe Conlon, Gonzalo Palma and Koenraad
  Schalm for very useful discussions. Work supported by the
  Netherlands Organization of Scientific Research (N.W.O.) under the
  VIDI and VICI programmes and by the Basque Government through grant
  BFI04.203 (K.S.) and project IT-357-07. We further acknowledge
  support by the Spanish Government through the Consolider-ingenio
  2010 Programme CPAN (CSD2007-00042) and project FPA2005-04823.}

\bigskip
\noindent\rule[0.0cm]{0.5in}{.1pt}%
\bigskip

\noindent
{\footnotesize
$^{2}$ \hspace{0.1cm}\parbox[t]{.46\textwidth}{[Refs. [\cite{Berg:2007wt}, \cite{Cicoli:2007xp}] suggest that string loop corrections to $\hat K$ scale as $(\text vol)^{-2/3}$ and would lead to $\delta < 10^{-6}$. We thank M. Cicoli for this remark.}}

\end{document}